# "Hi Sigma, do I have the Coronavirus?": Call for a New Artificial Intelligence Approach to Support Health Care Professionals Dealing With The COVID-19 Pandemic

Brian Subirana[Σ,Ω*], Ferran Hueto[Σ,Ω], Prithvi Rajasekaran[Σ], Jordi Laguarta[Σ], Susana Puig[σ], Josep Malvehy[σ], Oriol Mitja[ς], Antoni Trilla[σ], Carlos Iván Moreno [A], José Francisco Muñoz Valle[B], Ana Esther Mercado González[B,C], Barbara Vizmanos[B].Sanjay Sarma[Σ]

[Σ] Auto-ID Laboratory, Massachusetts Institute of Technology, Cambridge, MA, USA

[Ω] HES, Harvard University, Cambridge, MA, USA

[σ] Institut de Medicina i Dermatologia (ICMiD) Hospital Clínic de Barcelona, Universitat de Barcelona, Barcelona, Catalonia, Spain; Institut d'Investigacions biomèdiques August Pi i SUnyer (IDIBAPS), Barcelona, Catalonia, Spain; Centro de Investigación Biomédica en Red de Enfermedades Raras (CIBERER), Instituto de Salud Carlos III, Barcelona, Catalonia, Spain.

[ς] Infectious Diseases Department, Hospital Universitari Germans Trias i Pujol, Badalona, Catalonia, Spain; Fight AIDS and Infectious Disease Foundation.

[A] Coordinación General Académica, Universidad de Guadalajara, Guadalajara, Jalisco, México.

[B] Centro Universitario de Ciencias de la Salud; Universidad de Guadalajara, Guadalajara, Jalisco, México

[C] OPD Hospital Civil de Guadalajara, Guadalajara, Jalisco, México.

(∗) Corresponding Author: subirana@mit.edu; Phone:+1-617-952-3140; MIT Room 35-206, 77 Mass Ave., MA 02139

## Abstract

Just like your phone can detect what song is playing in crowded spaces, we show that Artificial Intelligence transfer learning algorithms trained on cough phone recordings results in diagnostic tests for COVID-19. To gain adoption by the health care community, we plan to validate our results in a clinical trial and three other venues in Mexico, Spain and the USA[1]. However, if we had data from other on-going clinical trials and volunteers, we may do much more. For example, for confirmed stay-at-home COVID-19 patients, a longitudinal audio test could be developed to determine contact-with-hospital recommendations, and for the most critical COVID-19 patients a success ratio forecast test, including patient clinical data, to prioritize ICU allocation. As a challenge to the engineering community and in the context of our clinical trial, the authors suggest distributing cough recordings daily, hoping other trials and crowdsourcing users will contribute more data. Previous approaches to complex AI tasks have either used a static dataset or were private efforts led by large corporations. All existing COVID-19 trials published also follow this paradigm. Instead, we suggest a novel open collective approach to large-scale real-time health care AI. We will be posting updates at https://opensigma.mit.edu. Our personal view is that our approach is the right one for large scale pandemics, and therefore is here to stay - will you join?

---

[1] The trial is focuses on Treatment of Mild Cases and Chemoprophylaxis of Contacts as Prevention of the COVID-19 Epidemic (CQ4COV19), planning to target 3040 participants which match the demographics were screening tests are more useful. Official title of the clinical trial: Antiviral Treatment of COVID-19 Confirmed Cases and Ring Chloroquine Chemoprevention in Close Contacts: a Cluster Randomized Clinical Trial.

## 1. An onset pre-screening test for COVID-19 is possible simply using cough phone recordings

Neither Siri, Alexa, nor Google can tell us if we have the Coronavirus despite the millions of expensive programing man-hours invested in them, nor can they support operational efficiency of related health care processes. Early detection of COVID-19 patients could drastically lower the spread of the disease[1,2], but it is not common practice yet because existing tests are not available at scale, require health care professionals' already limited time, are lengthy, and, are often wasted due to the lack of a reliable pre-screening test.

The goal of our research is weather such a test can be performed simply through phone-based cough recordings. In the past, speech recognition algorithms have been demonstrated to be able to perform various tasks related to cough detection.[3] In several cases involving different neurological conditions, researchers were able to develop machine learning algorithms that used free-flow speech to predict disease onset earlier than human experts, including psychosis with a sample size of less than 50,[4] and cognitive impairment with less than a thousand[5]. These studies effectively demonstrate the potential for Artificial Intelligence (AI) to achieve superhuman diagnosis capabilities, which we now believe can also be leveraged for COVID-19 diagnosis. Previous research shows audio recordings can be used to diagnose pneumonia[6], even from cheap cell phone recordings,[7] similar to the COVID-19 example shown in Figure 1. Dysphonia can be caused by and lead to the detection of inflammatory conditions such as allergies[8], infections[9], reflux[10], smoking[11] or trauma[12]. For emergency response teams, an effective AI first screening can perhaps be designed by combining voice recognition algorithms[13] with several questions

---

https://clinicaltrials.gov/ct2/show/NCT04304053?cond=covid&draw=2&rank=7. The other efforts are at Hospital Clínic de Barcelona, MIT and Universidad de Guadalajara.

selected by healthcare professionals. Imaging, if necessary, could be added for a third test, since it may distinguish coronavirus from other types of community acquired pneumonia.[14]

For such a test to be useful, one of the key requirements is that it can be developed quickly, e.g. without a lot of data if that can't be obtained quickly. Previous research has already shown deep learning can be useful through intra-data, low resource and unsupervised transfer learning on popular algorithms (VGG16, ResNet, EfficientNet, Universal Sentence Encoding, Word2Vec or GloVe). About 20.000 publications last year in this growing subdomain of AI means there are plenty of researchers and models to test.

Details of our implementation will be published elsewhere. After trying a few models, we modified a biologically inspired SOP, trained on regular speech. To do so, we started with a convolutional neural network trained on a regular speech dataset and then applied transfer learning. Transfer learning is a field of machine learning that focuses on improving the predictive power of an algorithm on a specific task by learning from a similar but distinct task. In formal terms, we can define a domain $D$ that is composed by a feature space $X$ and marginal distribution $P(X)$. A task for a given domain $D$ is defined as a label space $Y$ and predictive function $\eta$, which is based on a conditional distribution $P(Y|X)$. We can therefore define a source domain $D_S = \{X_S, P(X_S)\}$ with source task $T_S = \{Y_S, \eta_S\}$, and a target domain $D_T = \{X_T, P(X_T)\}$ with target task $T_T = \{Y_T, \eta_T\}$, where $X = \{x_1, \ldots, x_n\}, x_i \in X$ and $Y = \{y, \ldots, y_n\}, y_i \in Y$[15]. Transfer learning refers to the action of improving on the target task $T_T$'s predictive function $\eta_T$ from the applicable transferable information learned from $D_S$ and $T_S$, where $D_T \neq D_S$ and $T_T \neq T_S$. Note that transfer learning does not care about the predictive power of $\eta_S$ once $\eta_T$ has been updated, which is a field called multi-task learning[16]. Our implementation uses transfer learning from the domain of speech audio recordings $D_{Speech}$ to the target domain of COVID infected cough audio recordings $D_{Cough}$. Assuming audio recording and preprocessing is similar between both domains, we can define the specific case of $X_{Speech} = X_{Cough}$, but distinct label spaces $Y_{Speech} \neq Y_{Cough}$, which in the field of transfer learning is formally defined as heterogeneous transfer learning. This specific case of heterogeneous transfer learning requires for a domain adaption, which is to minimize the distance between marginal distributions ($P(X_{Speech}) \neq P(X_{Cough})$), and marginal distributions ($P(Y_{Speech}|X_{Speech}) \neq P(Y_{Cough}|X_{Cough})$). Feature based approaches (e.g. [17] [18] [19]) can be used to bridge this gap, where $P(X_{Speech}) \sim P(\xi(X_{Cough}))$ and $P(Y_{Speech}|X_{Speech}) \sim P(Y_{Cough}|\xi(X_{Cough}))$), where $\xi$ is a transformation function derived from features based transfer learning approaches. From this point on, this becomes a homogeneous transfer learning problem, which has been tackled in the past (and continues to be so) notably within instance-based (e.g.[20] [21]), symmetric and asymmetric feature-based (e.g.[22] [23]), parameter based (e.g. [24] [25]) and relational-based (e.g. [26] [27]) transfer learning approaches[28].

Specifically, within the field of deep learning, a popular approach for transfer learning is "off-the-shelf" feature generation[29], where a pretrained model from a source domain is used to generate a set of features from the task domain. These features are then processed using a shallow classifier (e.g. logistic regression, SVM, k-Nearest-Neighbors). This approach is especially utilized when the target domain $D_T$ has very few labeled examples, which is the case within our application, but tends to be surpassed by "fine-tuning" approaches when more data is available[30], by retraining specific layers from data within the target domain. We also tested this approach by pre-training DenseNet201[31] and ResNet50[32] architectures with samples from the speech dataset. We then used the output of the last layer of each of the models to generate features from our small set of cough recordings from COVID19 diagnosed and healthy patients (0.7 train, 0.3 test split).

We believe that the first stage in the transfer learning, i.e. that of selecting the non-cough domain, is what can have most impact in the results of the model. In fact, we evaluated the performance of four shallow machine learning classification algorithms (SVM, k-Nearest Neighbors, Random Forest, Logistic Regression) over a set of 5 cross validation test splits (see Figure 2). We then used Principal Component Analysis[33] to generate a visualization demonstrating the clustering between healthy and COVID-19 coughs respectively (see results on Figure 3 and 4). Our immediate next steps are to validate and grow our model as we get more data from our clinical trial.

In the next two sections we describe how we may do so collectively.

## 2. What else may speech-based diagnosis tell you about COVID-19?

To further validate and grow our model we aim to collect data on 150 patients and 3000 contacts as part of the clinical trial motioned in the abstract, which is focused on the COVID-19 onset. A second concurrent effort will look at Hospital Clínic de Barcelona patients focusses on symptomatic patients. We have also recently initiated data collection in Mexico and the USA. These four concurrent efforts may offer increased chances for transfer learning for our original screening test goal. We believe that if we broaden our sampling goals, we can also incorporate two news tests:

- For confirmed stay-at-home COVID-19 patients, a longitudinal audio test to determine contact-with-hospital recommendations.
- For the most critical COVID-19 patients, a success ratio to prioritize intensive care unit (ICU) allocation.

Thus, we have started a process to collect short audio-recording segments of 12 seconds for COVID-19 positive and control individuals, accepting them via WhatsApp, Web, Email or using an MIT developed Mobile App. Our underlying hypothesis is that several second-long voice samples can ultimately save millions of lives. We are focusing on four types of daily samples for each donating individual:

- Cough sounds
- The digits from 0 to 9
- The word "Ommmmmmm", with the "m" sound extending for the 12 seconds when possible

The rational for these choices follows from what is standard practice in the speech recognition community.[34] We conjecture that numeral pronunciation can signal type of voice patterns, while the extended "Om" sound may be detectably related to lung conditions.

Metadata useful to subject matter experts is often also helpful in developing AI algorithms (see supplementary table I for the one we will be collecting). It is difficult to say at the current time how important such metadata may be. In some cases, metadata not used by humans can yield superhuman results in AI algorithms. For example, unlike human specialists, AI can detect gender information from retina scans.[35]

Until we have a sizable and longitudinal sample database, we cannot be sure how many useful tests a COVID-19 detection algorithm can extract from sound signals. Indeed, it would be irresponsible to set upper or lower limits at this point. The data may be heavily biased towards one particular language, or the model may only apply to the onset of the disease – where the clinical trial is focusing on. As we discuss next, that is why we suggest an additional and different approach, one that depends on you the reader and many others.

## 3. Our personal view is that the rapid response to COVID-19-like pandemics calls for a new approach to Artificial Intelligence

Prevalent approaches to AI either enjoy a continuous pipeline of training data, such as the ones available to the FAANGs[2], or have the benefit of accessing a growing base of code, as in open source projects. The need for a new approach perhaps explains why AI has been largely absent from the infectious disease management debate, including all the current COVID-19 clinical trials we know of. Bill Gates did not even mention AI in his prophetic Ted Talk "The next outbreak? We are not ready"[3]. Others have advocated non-diagnostic uses of AI[36,37,38].

In contrast, we suggest a rapid development approach we call Sigma, where both data and code are shared real time. This approach means a novel collective effort by the Health and Engineering

---

[2] Facebook, Apple, Amazon, Netflix, Google (FAANGs).
[3] https://www.ted.com/talks/bill_gates_the_next_outbreak_we_re_not_ready

https://opensigma.mit.edu                                                                                                          Page 5 of 13 communities, where the first sets directions and provides patient samples in real time, while the second creates algorithms to improve infection management practices. Both groups will share samples, models and insights on a daily basis, further iterating sampling requirements, health practices and engineering efforts.

Sigma's objective is to set a research and engineering pipeline where daily samples received are immediately posted with an open source license for world-wide collaboration. The goal is to maximize the exposure of the samples to leverage talent worldwide as quickly as possible, with the three tests above as an initial challenge. AI algorithms have shown to be able to process unstructured data from various sources while incorporating cognitive data from experts. We will report elsewhere some of the legal hurdles we are encountering.

Sigma may become a template for future collaborations between engineering and medical professionals, but the urgent priority now is that other Hospitals join us in promoting the effort to feed the open sample database as quickly as possible. We are doing the groundwork so that MIT can accept samples donated by participating Hospitals and crowdsourcing individuals that are de-identified. This means that there is no reasonable basis to identify the recorded individual, and, that identifiers are not included as part of the metadata associated with the sample (such as names, telephone, age above 89, geographic subdivision of a certain size, etc.).

The three tests above are not the only urgent priority we may address collectively. As a byproduct of the data capturing process, we feel that voice interfaces can vastly increase ICU operational efficiency while minimizing infections. We expect other byproducts as a benefit of the Sigma approach. For example, the possibility for Physicians where the Pandemic has not yet arrived, to hear COVID-19 cough sounds or to see chest X-Rays for training. Or that the partial shutdown of the economy means vast idle computational and human resources may temporarily be donated to this effort.

Our suggested approach is unique because data is offered real time, as soon as is curated, allowing engineers to immediately test and improve candidate AI processing pipelines. Will you join?

**Supplementary Table I**

Sample Metadata included in the ongoing collaborative open project

- Age
- Gender
- Mother tongue (the one used in the samples)
- Region
- Days since onset of symptoms
- Comorbid conditions (most importantly the ones detected in COVID-19 research such as cardiovascular disease, pulmonary disease, diabetes or hypertension)[39]
- Concomitant medication
- In "in care" patients other information included as available:
    - RT-PCR for COVID-19
    - Blood test, including serologies for COVID-19
    - Thorax RX
    - Lung CT Scan
    - Need for ICU admission
    - Treatment

**Figure I**

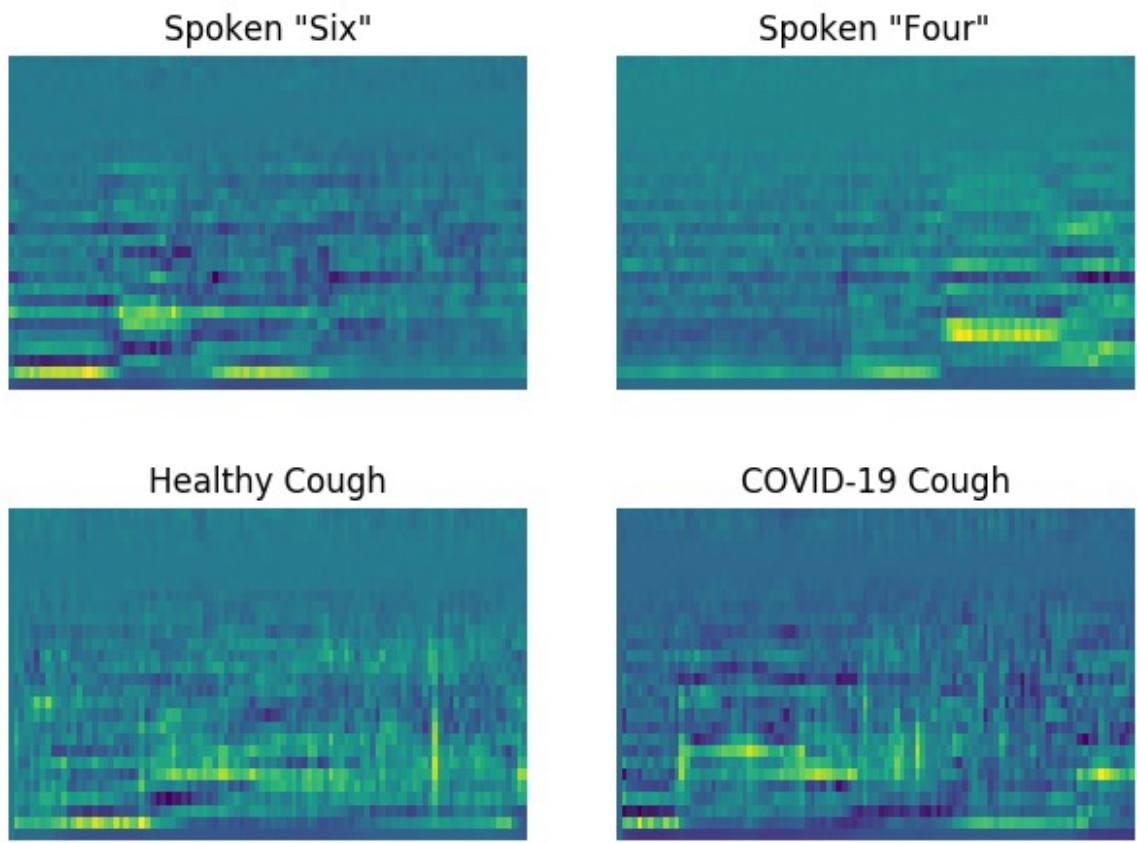

*Figure 1: Illustration of the differences between free speech, healthy cough and COVID-19 cough as recorded by a phone and received via WhatsApp. Digit samples were obtained from Google's Speech Commands Dataset[40], and stored in 1-channel 16.000 frame-rate, 2-byte sample width, non-compressed WAV files. The phone-recorded cough samples were sliced and converted from .OGG file format to match the same specifications. The sample plots where produced using Mel Frequency Cepstral Coefficients[41] on 0.99 second long raw audio files.*

**Figure II**

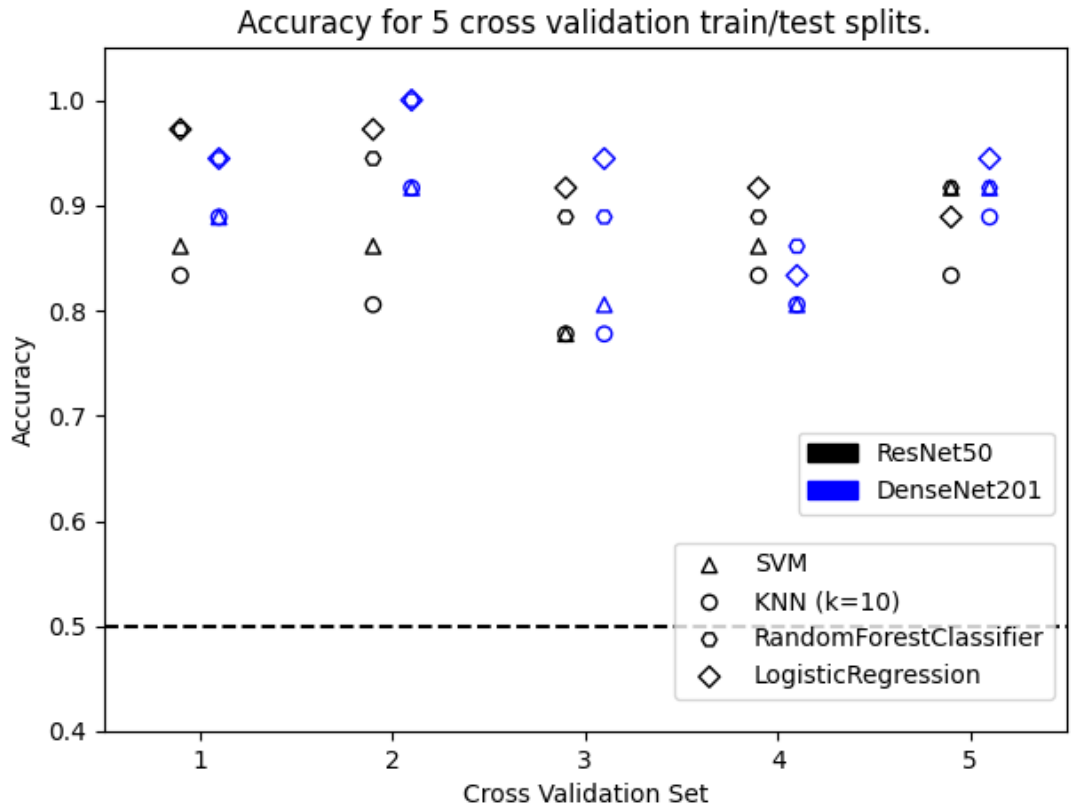

*Figure 2: Results from ResNet50 and DenseNet201 off-the-shelf features classified for COVID-19 diagnostic using four shallow machine learning algorithms (SVM, kNN, Random Forest and Logistic Regression)..*

**Figure III**

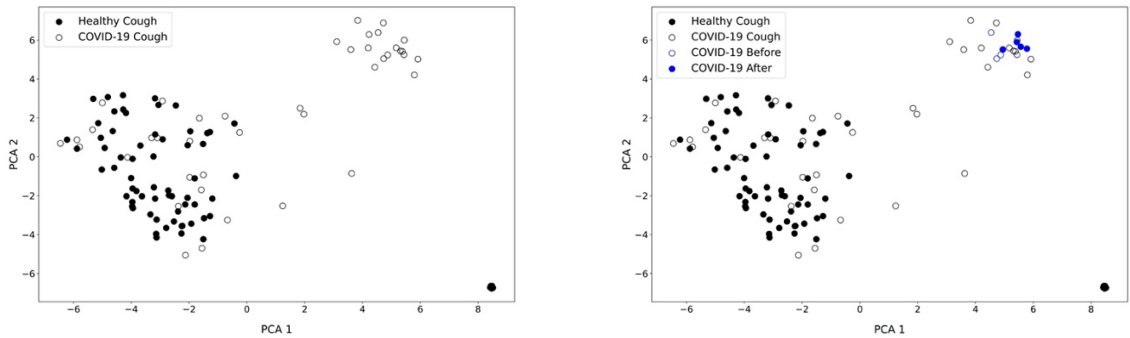

*Figure 3: Discriminating a COVID-19 cough from others is essential for the population of patients that don't exhibit any other significant synthoms. Using transfer learning from a spoken digit discrimination deep neural network, we have been able to automatically discriminate COVID-9 coughs from healthy coughs on a small dataset of under 200 samples, which includes such type of patients. Healthy and COVID-19 diagnosed tussication samples were run through the spoken digit model, 1024 features were extracted from the last convolutional layer and, for the purposes of the right plot, subsequently dimensionally reduced using Principal Component Analysis. The right plot demonstrates the addition of samples from the same individual at different stages of the disease.*

**Figure IV**

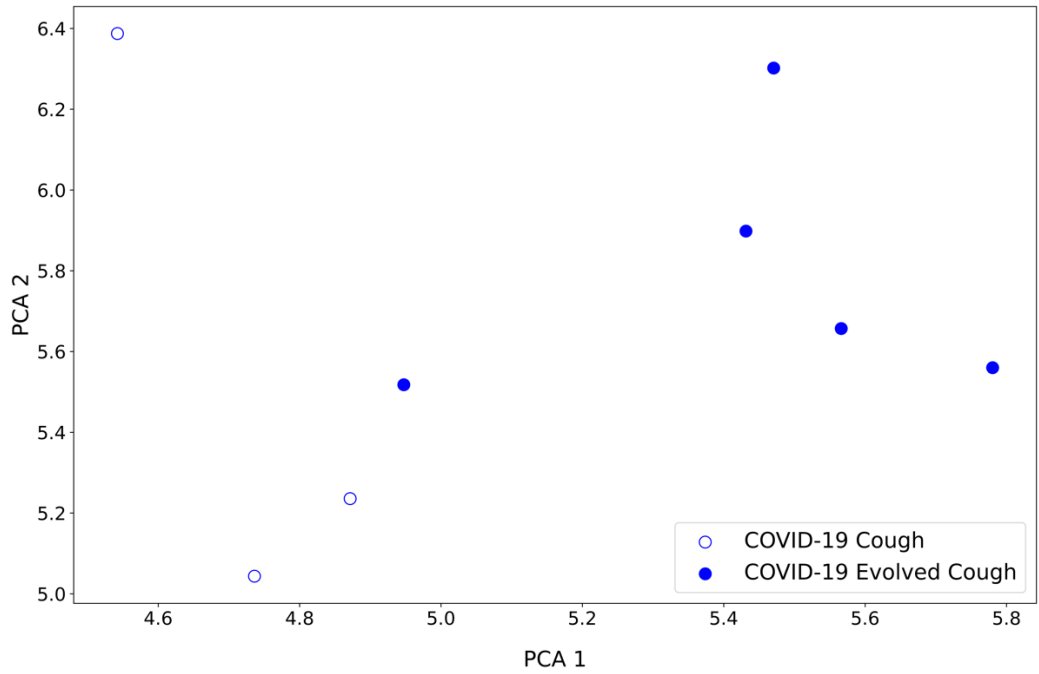

*Figure 4: Comparison of same "blue" subject as in Figure 3 with samples a few days a part.*

# References

[1] Xiao Y, Torok ME. Taking the right measures to control COVID-19. Lancet Infect Dis. 2020 Mar 5. pii: S1473-3099(20)30152-3.

[2] Wang D, Hu B, Hu C, Zhu F, Liu X, Zhang J, Wang B, Xiang H, Cheng Z, Xiong Y, Zhao Y, Li Y, Wang X, Peng Z. Clinical Characteristics of 138 Hospitalized Patients With 2019 Novel Coronavirus-Infected Pneumonia in Wuhan, China. JAMA. 2020 Feb 7. doi: 10.1001/jama.2020.1585

[3] Swarnkar, V., Abeyratne, U. R., Chang, A. B., Amrulloh, Y. A., Setyati, A., & Triasih, R. (2013). Automatic identification of wet and dry cough in pediatric patients with respiratory diseases. *Annals of biomedical engineering*, *41*(5), 1016-1028.

[4] Bedi, G., Carrillo, F., Cecchi, G. A., Slezak, D. F., Sigman, M., Mota, N. B., ... & Corcoran, C. M. (2015). Automated analysis of free speech predicts psychosis onset in high-risk youths. *npj Schizophrenia*, *1*, 15030.

[5] T. Alhanai, R. Au, and J. Glass, "Role-specific Language Models for Processing Neuropsychological Exams," *NAACL*, 2018; T. Alhanai, R. Au, and J. Glass, "Spoken Language Biomarkers for Detecting Cognitive Impairment," IEEE *ASRU*, 2017.

[6] Kosasih, K., Abeyratne, U.R., Swarnkar, V. and Triasih, R., 2014. Wavelet augmented cough analysis for rapid childhood pneumonia diagnosis. IEEE Transactions on Biomedical Engineering, 62(4), pp.1185-1194.

[7] Song, I., 2015, July. Diagnosis of pneumonia from sounds collected using low cost cell phones. In 2015 International joint conference on neural networks (IJCNN) (pp. 1-8). IEEE.

[8] Susanna Simberg, et al, 2009. Vocal Symptoms and Allergy–A Pilot Study. *Journal of Voice,* 23(1), pp. 136-139.

[9] Roberts, J. E. & Zeisel, S. A., 1969. Effect of chronic otitis media on language and speech development. *Pediatrics,* 43(5), pp. 833-839.

[10] James A. Koufman, et al 2002. Laryngopharyngeal Reflux: Position Statement of the Committee on Speech, Voice, and Swallowing Disorders of the American Academy of Otolaryngology-Head and Neck Surgery. *Otolaryngology–Head and Neck Surgery,* 127(1), pp. 32-35.

[11] David Sorensen, et al 1982. Cigarette smoking and voice fundamental frequency. *Journal of Communication Disorders,* 15(2), pp. 135-144.

[12] Terese Finitzo, et al 1987. Spasmodic Dysphonia Subsequent to Head Trauma. *Arch Otolaryngol Head Neck Surg,* 113(10), pp. 1107-1110.

[13] Swarnkar, V., Abeyratne, U. R., Chang, A. B., Amrulloh, Y. A., Setyati, A., & Triasih, R. (2013). Automatic identification of wet and dry cough in pediatric patients with respiratory diseases. *Annals of biomedical engineering*, *41*(5), 1016-1028.

[14] Xu, X., Jiang, X., Ma, C., Du, P., Li, X., Lv, S., ... & Li, Y. (2020). Deep Learning System to Screen Coronavirus Disease 2019 Pneumonia. *arXiv preprint arXiv:2002.09334*.

[15] Pan SJ, Yang Q. A survey on transfer learning. IEEE Trans Knowl Data Eng. 2010;22(10):1345–59

[16] Evgeniou, T. and Pontil, M., 2004, August. Regularized multi--task learning. In Proceedings of the tenth ACM SIGKDD international conference on Knowledge discovery and data mining (pp. 109-117).

[17] Zhao, J., Shetty, S. and Pan, J.W., 2017, October. Feature-based transfer learning for network security. In MILCOM 2017-2017 IEEE Military Communications Conference (MILCOM) (pp. 17-22). IEEE.

[18] Tzeng, E., Hoffman, J., Darrell, T. and Saenko, K., 2015. Simultaneous deep transfer across domains and tasks. In Proceedings of the IEEE International Conference on Computer Vision (pp. 4068-4076).

[19] Han, T., Liu, C., Yang, W. and Jiang, D., 2019. Deep transfer network with joint distribution adaptation: A new intelligent fault diagnosis framework for industry application. ISA transactions.

[20] Lin, D., An, X. and Zhang, J., 2013. Double-bootstrapping source data selection for instance-based transfer learning. Pattern Recognition Letters, 34(11), pp.1279-1285.

[21] Chattopadhyay R, Ye J, Panchanathan S, Fan W, Davidson I. Multi-source domain adaptation and its application to early detection of fatigue. ACM Trans Knowl Dis Data (Best of SIGKDD 2011 TKDD Homepage archive) 2011; 6(4)

[22] Vincent, P., Larochelle, H., Lajoie, I., Bengio, Y. and Manzagol, P.A., 2010. Stacked denoising autoencoders: Learning useful representations in a deep network with a local denoising criterion. Journal of machine learning research, 11(Dec), pp.3371-3408.

[23] Kandemir, M., 2015, June. Asymmetric transfer learning with deep gaussian processes. In International Conference on Machine Learning (pp. 730-738).

[24] Tommasi T, Caputo B. The more you know, the less you learn: from knowledge transfer to one-shot learning of object categories. BMVC. 2009;1–11.

[25] Tommasi T, Orabona F, Caputo B. Safety in numbers: learning categories from few examples with multi model knowledge transfer. IEEE Conf Comput Vision Pattern Recog. 2010;2010:3081–8.

[26] Wang, D., Li, Y., Lin, Y. and Zhuang, Y., 2016, March. Relational knowledge transfer for zero-shot learning. In Thirtieth AAAI Conference on Artificial Intelligence.

[27] Mihalkova, L. and Mooney, R.J., 2009, June. Transfer learning from minimal target data by mapping across relational domains. In Twenty-First International Joint Conference on Artificial Intelligence.

[28] Weiss, K., Khoshgoftaar, T.M. and Wang, D., 2016. A survey of transfer learning. Journal of Big data, 3(1), p.9.

[29] Sharif Razavian, A., Azizpour, H., Sullivan, J. and Carlsson, S., 2014. CNN features off-the-shelf: an astounding baseline for recognition. In Proceedings of the IEEE conference on computer vision and pattern recognition workshops (pp. 806-813).

[30] Yosinski, J., Clune, J., Bengio, Y. and Lipson, H., 2014. How transferable are features in deep neural networks?. In Advances in neural information processing systems (pp. 3320-3328).

[31] Iandola, F., Moskewicz, M., Karayev, S., Girshick, R., Darrell, T. and Keutzer, K., 2014. Densenet: Implementing efficient convnet descriptor pyramids. arXiv preprint arXiv:1404.1869.

[32] He, K., Zhang, X., Ren, S. and Sun, J., 2016, October. Identity mappings in deep residual networks. In European conference on computer vision (pp. 630-645). Springer, Cham.

[33] Zou, H., Hastie, T. and Tibshirani, R., 2006. Sparse principal component analysis. *Journal of computational and graphical statistics*, *15*(2), pp.265-286.

[34] See for example the 2020 ADReSS Challenge (Alzheimer's Dementia Recognition through Spontaneous Speech) at http://www.homepages.ed.ac.uk/sluzfil/ADReSS/.

[35] Poplin, R., Varadarajan, A.V., Blumer, K., Liu, Y., McConnell, M.V., Corrado, G.S., Peng, L. and Webster, D.R., 2018. Prediction of cardiovascular risk factors from retinal fundus photographs via deep learning. *Nature Biomedical Engineering*, *2*(3), p.158.

[36] Allam, Zaheer, and David S. Jones. "On the Coronavirus (COVID-19) Outbreak and the Smart City Network: Universal Data Sharing Standards Coupled with Artificial Intelligence (AI) to Benefit Urban Health Monitoring and Management." *Healthcare*. Vol. 8. No. 1. Multidisciplinary Digital Publishing Institute, 2020.

[37] Rao, A., & Vazquez, J. (n.d.). Identification of COVID-19 Can be Quicker through Artificial Intelligence framework using a Mobile Phone-Based Survey in the Populations when Cities/Towns Are Under Quarantine. *Infection Control & Hospital Epidemiology,* 1-18. doi:10.1017/ice.2020.61

[38] JOHNSTONE, Syren. A Viral Warning for Change. The Wuhan Coronavirus Versus the Red Cross: Better Solutions Via Blockchain and Artificial Intelligence. *The Wuhan Coronavirus Versus the Red Cross: Better Solutions Via Blockchain and Artificial Intelligence (February 3, 2020)*, 2020.

[39] Wang, Dawei, et al. "Clinical characteristics of 138 hospitalized patients with 2019 novel coronavirus–infected pneumonia in Wuhan, China." *Jama* (2020)

[40] Warden, P., 2018. Speech commands: A dataset for limited-vocabulary speech recognition. *arXiv preprint arXiv:1804.03209*.

[41] Molau, S., Pitz, M., Schluter, R. and Ney, H., 2001, May. Computing mel-frequency cepstral coefficients on the power spectrum. In *2001 IEEE International Conference on Acoustics, Speech, and Signal Processing. Proceedings (Cat. No. 01CH37221)* (Vol. 1, pp. 73-76). IEEE.